\documentclass[runningheads]{llncs}
\usepackage{graphicx}
\usepackage{bm}
\usepackage{amssymb}
\usepackage{booktabs}
\begin{document}
\title{THUIR@COLIEE-2020: Leveraging Semantic Understanding and Exact Matching for Legal Case Retrieval and Entailment\thanks{This work is supported by the National Key Research and Development Program of China (2018YFC0831700), Natural Science Foundation of China (Grant No. 61732008, 61532011), Beijing Academy of Artificial Intelligence (BAAI)and Tsinghua University Guoqiang Research Institute.}}
\titlerunning{THUIR@COLIEE2020}
%
\author{
Yunqiu Shao\and
Bulou Liu\and
Jiaxin Mao\and
Yiqun Liu\thanks{Corresponding Author}\and
Min Zhang\and
Shaoping Ma
}
\authorrunning{Y. Shao et al.}
%
\institute{Department of Computer Science and Technology, Institute for Artificial Intelligence,
Beijing National Research Center for Information Science and Technology,
Tsinghua University, Beijing 100084, China \\
\email{yiqunliu@tsinghua.edu.cn}
}

\pagestyle{empty}  
\thispagestyle{empty} 

\maketitle              
\begin{abstract}
In this paper, we present our methodologies for tackling the challenges of legal case retrieval and entailment in the Competition on Legal Information Extraction/Entailment 2020 (COLIEE-2020). We participated in the two case law tasks, i.e., the legal case retrieval task and the legal case entailment task. Task 1 (the retrieval task) aims to automatically identify supporting cases from the case law corpus given a new case, and Task 2 (the entailment task) to identify specific paragraphs that entail the decision of a new case in a relevant case. In both tasks, we employed the neural models for semantic understanding and the traditional retrieval models for exact matching. As a result, our team (``TLIR'') ranked 2nd among all of the teams in Task 1 and 3rd among teams in Task 2. Experimental results suggest that combing models of semantic understanding and exact matching benefits the legal case retrieval task while the legal case entailment task relies more on semantic understanding. 


\end{abstract}
\section{Introduction}
Precedents are primary legal materials in the common law systems. For instance, retrieving and reviewing supporting precedents is fundamental for a lawyer who is preparing the legal reasoning in the countries that follows the common law system (e.g., USA, Canada). However, it takes a great number of efforts of legal practitioners to search for relevant cases and extract the entailment parts manually with the rapid growth of digitalized legal documents. Therefore, an efficient legal assistant system that would alleviate the heavy document work has drawn increasing attention in the field of AI \& Law. 

Being held since 2014, the Competition on Legal Information Extraction / Entailment (COLIEE) is a series of evaluation competitions aiming to develop techniques for information retrieval and entailment using legal texts~\cite{juliano2019coliee}. In COLIEE-2020\footnote{https://sites.ualberta.ca/\~rabelo/COLIEE2020/}, there are four tasks in total, among which, Task 1 \& 2 are about the case law while Task 3 \& Task 4 are about the statute law. Our team participated in the case law tasks (Task 1 \& 2) this year. In previous studies, a variety of NLP and IR strategies have been developed and applied in these tasks. For example, Vu et al.~\cite{tran2019building} proposed a phrase scoring model and developed a summarization-based model for the legal case retrieval task. The application of the pre-trained language model BERT~\cite{devlin2018bert} has also been investigated. To deal with the long case documents, Rossi et al.~\cite{Julien2019legal} attempted to generate the summary of the case via unsupervised auto summarization algorithms (e.g., TextRank~\cite{mihalcea2004textrank}) first and feed them into the BERT model. On the other hand, Shao et al.~\cite{shaobert} proposed to model paragraph-level interactions to avoid information loss in the summarization process when dealing with long documents.

In this paper, we introduce our approaches to completing Task 1 and Task 2. Our models consider both exact matching and semantic understanding as well as the combination of them. To be specific, we utilize the word-entity duet framework~\cite{xiong2017word} (exact matching) and the BERT-PLI model~\cite{shaobert} (semantic understanding) in the first two runs in Task 1. Moreover, we extract features from these two methods and combine them through the learning to rank strategy (combination). In Task 2, we fine-tune BERT with symmetric and asymmetric truncation in the first two runs and combine the representation features with others, including the exact matching features and the meta-features. As a result, our team ranked 2nd among all of the teams in Task 1 and 3rd among teams in Task 2.

\section{Task Description}
\subsection{Task 1: Legal Case Retrieval}
Task 1 is a legal case retrieval task, which involves reading a new case (query case $Q$) and extracting supporting cases ($D=\{d_1, d_2, ..., d_n\}$) from the provided case law corpus. The ``supporting case'' (or ``noticed case'') is a legal term that denotes the precedent can support the decision of a query case, and thus in our paper, we consider it as the relevant case in the scenario of the legal case retrieval. 


\subsection{Task 2: Legal Case Entailment}
Task 2 is a legal case entailment task, which involves identifying the paragraphs that entail the decision of a new case. Formally, given a decision fragment of a new case (denoted as $q$) and a relevant case composed of paragraphs (denoted as $R=\{p_1, p_2, ..., p_n\}$), the task is to find all of the $p_i$ satisfying $entail(q, p_i)$. Different from the retrieval task, a relevant paragraph does not necessarily entail the query fragment. 


\subsection{Dataset}
Data of Task 1 and Task 2 are both drawn from an existing collection of predominantly Federal Court of Canada case law. Table~\ref{dataset} gives a statistical summary of the dataset. In Task 1, the training and testing set consist of 520 and 130 query cases respectively, and each query case is provided with 200 candidate cases. In Task 2, the training and testing set have 325 and 100 base cases respectively, along with the corresponding decision fragment extracted from the base case. Besides, regarding each base case, a relevant precedent is given in the form of a paragraph list. The golden labels of the training set are provided in both tasks. Participants are required to submit the retrieved results on the testing set for evaluation. 

\begin{table}[]
\caption{Summary of datasets in COLIEE 2020 Task 1 and Task 2.}
\centering
\begin{tabular}{lrr}
\toprule
\textbf{Task 1}                  & \textbf{Train} & \textbf{Test}\\ 
\midrule
\# query case                   & 520   & 130   \\
\# candidate cases / query      & 200   & 200   \\
\# noticed cases / query        & 5.15  & --    \\ \midrule
\textbf{Task 2}                 & \textbf{Train} & \textbf{Test}\\ 
\midrule
\# query case                   & 325   & 100   \\
\# candidate paragraphs / query & 35.52 & 36.72 \\
\# entailing paragraphs / query & 1.15  & --    \\ \bottomrule
\end{tabular}
\label{dataset}
\end{table}

\subsection{Evaluation Measure}
Both tasks are evaluated with micro-average of precision, recall, F1 score, where

\begin{equation}
    Precision = \frac{Q_{TP}}{Q_{TP+FP}}
\end{equation}

\begin{equation}
    Recall = \frac{Q_{TP}}{Q_{TP+FN}}
\end{equation}

\begin{equation}
    F1 = \frac{2 \cdot Precision \cdot Recall}{Precision+Recall}
\end{equation}
where $Q_{TP}$ denotes the number of \textbf{correctly} retrieved cases (paragraphs) for all queries, $Q_{TP+FP}$ is the number of retrieved cases (paragraphs) for all queries, and $Q_{TP+FN}$ is the number of labels for all queries.

\section{Methods}
\subsection{Task 1: Legal Case Retrieval}
\subsubsection{Run1 - Word-entity Duet.}
Traditional retrieval models are mostly calculated by term-level exact matching based on bag-of-words representations. Beyond bag-of-words, bag-of-entities (or bag-of-concepts) have been investigated in various scenarios, such as the ad-hoc text retrieval~\cite{xiong2017word,chen2016empirical} and medical retrieval~\cite{lee2018seed}. Shao et al.~\cite{shaobert} has pointed out that traditional bag-of-words IR models have competitive performances in legal case retrieval. Therefore, we would like to further study the exact-matching-based models and expand the bag-of-words representation to a mixed one. In particular, we employ the word-entity duet framework~\cite{xiong2017word} in the first run. Each case document is modeled with word-based and entity-based representations. Then ranking features are generated to incorporate information from the word space, the entity space, and the cross-space connections. Due to the lack of public knowledge based in the Canadian legal domain, we simply utilize a general NER tool\footnote{https://www.nltk.org/} to extract entities in the case texts. Through our pilot study, we find that the model performs better if we dismiss the term frequency in the entity-based features. Therefore, we follow this setting in this run.

Within the word-entity duet framework, a query case can interact with a candidate case document in different ways, including from query words to document words (Qw-Dw), from query entities to document words (Qe-Dw), and from query words to document entities (Qw-De). The interaction features are calculated by matching-based retrieval models, as shown in Table~\ref{duet_features}. In total, 11-dimensional features are utilized for ranking. 



\begin{table}[]
\caption{Ranking features used in the word-entity duet model.}
\centering
\begin{tabular}{lr}
\toprule
\textbf{Interaction} & \textbf{Feature Description } \\ 
\midrule
Qw-Dw & BM25                        \\
Qw-Dw & TF-IDF                      \\
Qw-Dw & LM                          \\
Qw-Dw & Lm with JM smoothing        \\
Qw-Dw & Lm with Dirichlet smoothing \\
Qw-Dw & Lm with two-way smoothing   \\
\midrule
Qe-Dw & BM25                        \\
Qe-Dw & TF-IDF                      \\
Qe-Dw & Lm with Dirichlet Smoothing \\
\midrule
Qw-De & TF-IDF                      \\
Qw-De & Lm-Dirichlet                \\ 
\bottomrule
\end{tabular}
\label{duet_features}
\end{table}

Based on the selected features, the relevance score between the query case and the candidate case is calculated by, 
\begin{equation}
    f(q,d)=g_a(W^T_{duet} \cdot v_{duet} + b_{duet})
\end{equation}
, where $W_{duet}$ and $b_{duet}$ denote the weight vector and bias respectively, and $g_a(\cdot)$ is the \textit{sigmoid} function. 

Following Xiong et al.~\cite{xiong2017word}, the model is trained by optimizing the pairwise hinge loss:
\begin{equation}
     L_{hinge}(q,D)=\sum_{d \in D^+}\sum_{d' \in D^-}max(0, 1-f(q,d)+f(q,d'))
\end{equation}
, where $D^+$ and $D^-$ are the set of relevant and irrelevant candidate cases. The loss function can be optimized using back-propagation in the neural network. 

We consider it as a ranking problem in this run and return the top-5 ranked documents as the relevant cases. Before generating the rank features, we pre-process the documents by removing the punctuation and stop words via the NLTK tool. Parameters of the retrieval models (e.g. LM, BM25, etc.) in this run are kept as default. The learning rate of pairwise training is set as $1e-4$ without weight decay. The word-entity duet model is trained on the split training data for no more than 20 epochs and selected according to the F1 score on the split validation set. The division of the dataset would be detailed discussed in Section~\ref{sec:exp}.

\subsubsection{Run2 - BERT-PLI.} Shao et al.~\cite{shaobert} attempt to tackle the challenges caused by long and complex documents by modeling paragraph-level interactions and thus propose the \textbf{BERT-PLI} for legal case retrieval. Compared to traditional retrieval models, it has a better semantic understanding ability. Meanwhile, it can utilize the entire legal case document that is too long for the neural models developed for ad-hoc text retrieval to deal with. In COLIEE-2020, there still exist these challenges in the legal case retrieval task, including the extremely long documents, the complex definition of legal relevance, and the limited scale of training data. Therefore, we consider this task as a binary classification one and employ BERT-PLI in our second run.

\begin{figure}
    \centering
    \includegraphics[width=0.9\columnwidth]{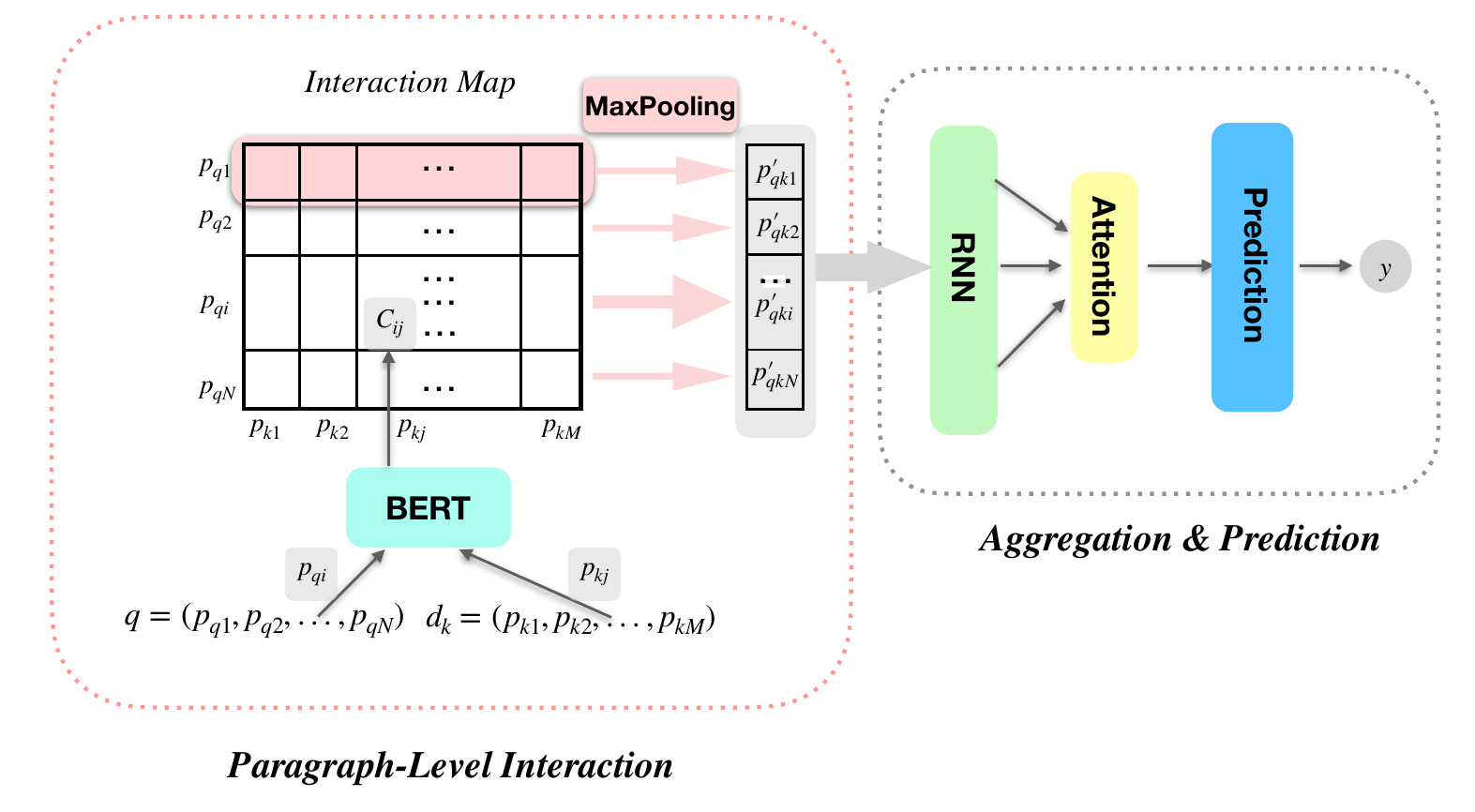}
    \caption{An illustration of BERT-PLI~\cite{shaobert}.}
    \label{fig:bertpli}
\end{figure}

As illustrated in Figure~\ref{fig:bertpli}, for each input pair of query case and candidate case, $(q, d_k)$, we first represent the case document by paragraphs, denoted by $q=(p_{q1}, p_{q2}, ..., p_{qN})$ and $d_k = (p_{k1}, p_{k2}, ..., p_{kM})$ respectively. In total, we obtain $N \times M $ pairs of paragraphs. Then we apply BERT to infer the semantic relationship of each paragraph pair, e.g., $(p_{qi}, p_{kj})$, and utilize the final hidden vector of [CLS] token as the representation of the corresponding paragraph pair. Once the interaction map through BERT inference is constructed, the \textit{maxpooling} strategy is used to capture the strongest matching signals for each paragraph of the query case. The we utilize a forward RNN structure combined with the attention mechanism to further encode and aggregate the paragraph-based representation sequence, $\bm{p}'_{qk} = [\bm{p}'_{qk1}, \bm{p}'_{qk2}, \ldots, \bm{p}'_{qkN}]$, into a document-based representation, $\bm{d}_{qk}$. As for prediction, we pass the representation $\bm{d}_{qk}$ through a fully connected layer followed by a \textit{softmax} function and optimize the cross-entropy loss during the training process. 

Following Shao et al.~\cite{shaobert}, we adopt a cascade framework considering the computational cost. Specifically, we rank all of the candidate documents ahead according to the scores given by bi-gram LMIR (\textbf{L}anguage \textbf{M}odel for \textbf{I}nformation \textbf{R}etrieval)~\cite{song1999general} with a linear smoothing referring to prior work~\cite{shao2019thuir}, and select the top-30 case as the candidates that would be further classified by BERT-PLI. 

In our run, the uncased based BERT~\cite{devlin2018bert} version\footnote{https://github.com/google-research/bert} is employed. Moreover, we do not update the parameters of BERT during the training process of BERT-PLI but utilize the parameters~\cite{shaobert} that were fine-tuned with a legal paragraph dataset. The model settings as well as the training parameters all following the prior work~\cite{shaobert}, in which $N=54$, $M=40$, $lr=1e-4$ with a weight decay of $1e-6$. In practice, we consider LSTM and GRU respectively as the RNN component and the results in the validation set suggest that GRU achieves better performance. Therefore, in the submitted run, a one-layer GRU with 256 hidden units is used. We train BERT-PLI on the split training data for no more than 60 epochs and select the best one according to the F1 score on the split validation set.

\subsubsection{Run3 - Combination.} Our first run is focused on the exact matching between two case documents based on traditional bag-of-words and bag-of-entities models, while our second run pays more attention to the semantic understanding through neural models. In this run, we attempt to combine the models of exact matching and semantic understanding. To be specific, we extract features of both kinds of models and employ learning to rank algorithms~\cite{liu2011learning} to rank the candidate cases. 

As for the features of exact matching, we use the 11-dimensional features listed in Table~\ref{duet_features}. Besides, we apply the SDR (\textbf{S}eed-driven \textbf{D}ocument \textbf{R}anking)~\cite{lee2018seed}, which is developed for the systematic review in medical retrieval, to the legal case retrieval task, and consider the similarity scores (calculated based on words and entities, respectively) as the additional two exact matching features. On the other hand, we extract the output vector (two-dimensional) of the \textit{softmax} function in BERT-PLI as the features of semantic understanding. Besides, we use the first paragraph of the query and a candidate case as the input of BERT to fine-tune a sentence pair classification task~\footnote{The fine-tuning process is described in detail in Section~\ref{subsec:task2}, run1} and extract the vector (two-dimensional) given by \textit{softmax} function as additional features. In total, we obtain 13-dimensional (11+2) features based on exact matching and 4-dimensional features (2+2) based on semantic understanding. Note that BERT-PLI is only calculated on the top-30 candidate cases ranked by LMIR, so in this run, only the features of these top-30 candidates are considered. 

We employ the RankSVM\footnote{https://www.cs.cornell.edu/people/tj/svm\_light/svm\_rank.html} to re-rank the top-30 candidates based on the 17-dimensional features. Specifically, we tune the parameter ``-c'' according to the F1 measure on the validation set and set $c=20$ in this run. For each query, we consider the cases of non-negative predicted scores as the relevant ones. Meanwhile, if a query case has less than 3 relevant cases, we rank the remaining case based on the predicted scores and append the highly ranked ones to the result list until the query has 3 relevant cases. 

\subsection{Task 2: Legal Case Entailment}
\label{subsec:task2}
\subsubsection{Run1 - BERT Fine-tuning.} 
\begin{figure}
    \centering
    \includegraphics[width=0.5\columnwidth]{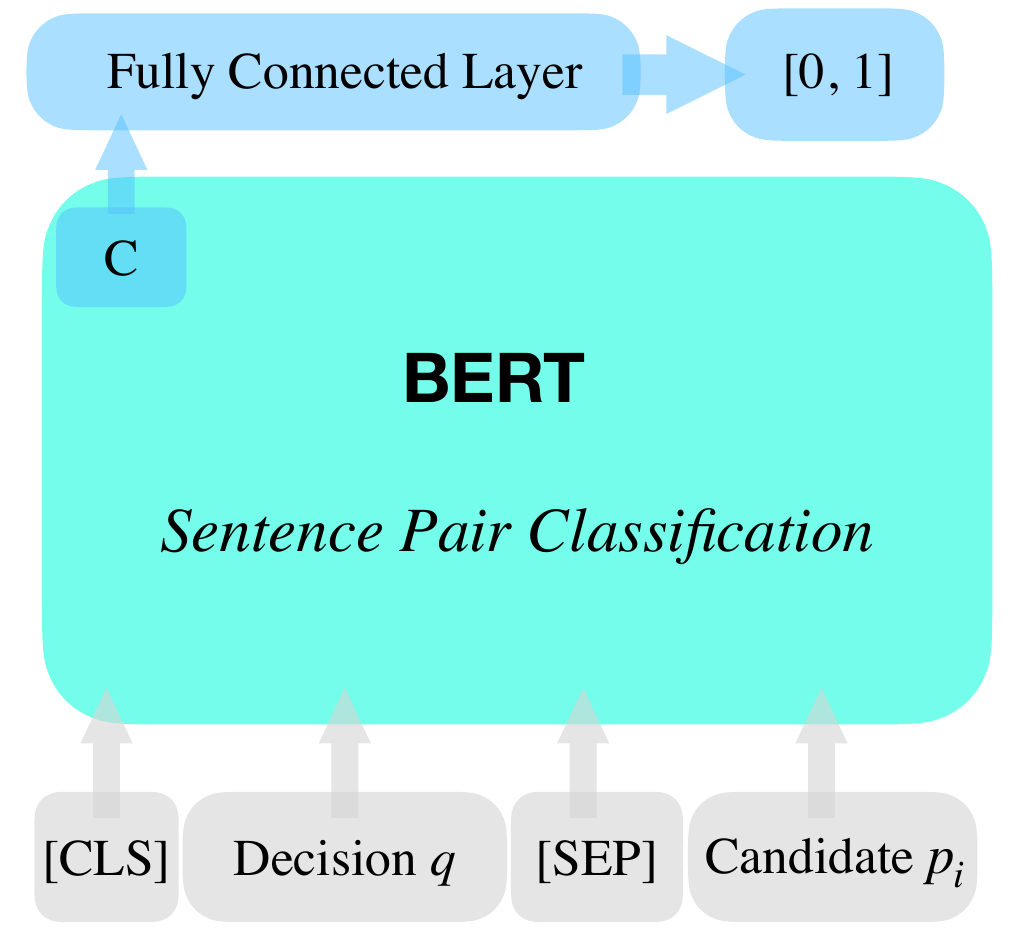}
    \caption{Fine-tuning a sentence pair classification task.}
    \label{fig:finetune}
\end{figure}

We fine-tune the BERT~\cite{devlin2018bert} with a downstream sentence pair classification task. As illustrated in Figure~\ref{fig:finetune}, the input pair of text is composed of the decision fragment in the query case and a candidate paragraph in the relevant case, denoted as $(q, p_i)$, where $p_i$ denotes the $i$-th paragraph in the corresponding relevant case. The final hidden state vector of [CLS] is then fed into a fully-connected layer for binary classification. 

In this run, we make use of all of the candidate paragraphs in the given relevant case to construct the input text pair. If the total input tokens exceed the length limitation (512), the texts are truncated symmetrically. The model is training on the split training set by optimizing the \textit{cross-entropy} loss. We update all of the parameters in an end-to-end way. The model is trained for no more than 5 epochs with $lr=1e-5$ and selected according to the F1 measure on the validation set. The division of the dataset would be described in Section~\ref{sec:exp}. As a result, we employ the one that trained for 4 epochs as the submitted run. 

\subsubsection{Run2 - BERT Fine-tuning with Asymmetric Truncation.}
The main difference from ``Run1'' is in that we truncate the text asymmetrically if the input tokens exceed the length limitation. In the training data, we observe that most of the decision fragment is no longer than 128 words while the paragraphs in the relevant case differ a lot. In the case where the candidate paragraph is rather long, symmetric truncation might lead to severe information loss in the decision fragment. Therefore, we are inspired to keep the tokens of decision fragment identical in each text pair corresponding to the query case.

In detail, we limit the tokens of decision fragment to 128, in which case, we truncate the decision fragment $q$ ahead if it exceeds the limit. Then, we construct the input text pair combining with the candidate paragraphs, and we only truncate the tokens in the candidate paragraph $p_i$ if the total length of the text pair exceeds 512 tokens. The model structure as well as the training process is the same as that in ``Run1''. With Asymmetric truncation, the training process converges faster and we select the model trained for 1 epoch as the submitted run. 

\subsubsection{Run3 - Combination.} In this run, we also attempt to combine models of semantic understanding and exact matching along with some meta-information. Different from the retrieval task, the entailment task emphasizes the semantic understanding since the relevant/similar paragraphs might not entail the decision fragment. Therefore, we pay more attention to the features of semantic understanding models. To be specific, we extract the output vector of the fully-connected layer of ``Run 1'' and ``Run 2'' (4-dimensional in total). As for the features of exact matching, we only consider the BM25 scores (1-dimensional) implemented by gensim~\footnote{https://radimrehurek.com/gensim/} with default parameters. Moreover, considering the legal document is somehow semi-supervised, we attempt to utilize the structure features including the position ID and the length of the paragraph (2-dimensional in total). To sum up, we obtain 7-dimensional features in this run and then RankSVM is employed based on them. We consider the paragraphs with non-negative predicted scores as the entailment paragraph, and if all of the predicted scores are negative, we select the top-ranked one as the result. Similarly, we tune the parameter ``-c'' according to the F1 measure on the validation set, and set $c=1$ in this run. 

\section{Experiments and Results}
\label{sec:exp}
For both tasks, we divide the training data into two separate parts randomly. 20\% of queries as well as all of their candidates are treated as the validation set, and the remaining queries along with the candidates are utilized for training. We maintain the same division of training and validation set across all of the runs in the corresponding task. 

\subsection{Task1: Legal Case Retrieval}
\begin{table}[]
    \centering
    \caption{Results on the test set of Task 1.}
    \begin{tabular}{llrr}
    \toprule
    \textbf{Team ID} & \textbf{Run ID} & \textbf{F1} & \textbf{Rank} \\
    \midrule
    \textbf{TLIR} & \textbf{Run 3} & \textbf{0.6682} & \textbf{3} \\
    TLIR & Run 2 & 0.6379 & 6 \\
    TLIR & Run 1 & 0.5148 & 12 \\
    \midrule
    \textbf{cyber} & \textbf{Run 2} & \textbf{0.6774} & \textbf{1} \\
    cyber & Run 3 & 0.6768 & 2 \\
    cyber & Run 1 & 0.6503 & 4 \\
    \midrule
    JNLP & BMW25 & 0.6397 & 5 \\
    \bottomrule
    \end{tabular}
    \label{tab:task1}
\end{table}

Table~\ref{tab:task1} gives the results on the test data in Task 1, the legal case retrieval task\footnote{For simplicity, we only list parts of the runs here that we would like to further interpret}. Among our submitted runs (``TLIR''), Run 3, which combines the features of semantic understanding and exact matching, performs the best and outperforms the other single models in a nontrivial scale. This result suggests that the models of semantic understanding and the ones of exact matching focus on different aspects of legal documents, and thus can be complementary in the legal case retrieval task. Comparing the result of BERT-PLI (Run 2) with that of Duet (Run 1), the development of semantic neural models enhance the performance in the legal case retrieval a lot. We assume that this task calls for the ability of semantic understanding beyond the exact keywords matching due to the complexity of relevance definition in the legal domain. 

The results are ranked according to the F1 score, and the top-2 runs are both from the team ``cyber''. We note that the difference between our method and the top-1 run in F1 is less than 0.01. Considering the limited data scale (130 test queries in total), we believe that our third method achieves rather close performance to the cyber's in this task.

\subsection{Task2: Legal Case Entailment}
\begin{table}[]
    \centering
    \caption{Results on the test set of Task 2.}
    \begin{tabular}{llrr}
    \toprule
    \textbf{Team ID} & \textbf{Run ID} & \textbf{F1} & \textbf{Rank} \\
    \midrule
    \textbf{TLIR} & \textbf{Run 2} & \textbf{0.6154} & \textbf{4} \\
    TLIR & Run 3 & 0.5495 & 13 \\
    TLIR & Run 1 & 0.5428 & 14 \\
    \midrule
    \textbf{JNLP} & \textbf{BMWT} & \textbf{0.6753} & \textbf{1} \\
    JNLP & BMW & 0.6222 & 2 \\
    taxi & XGBaft & 0.6180 & 3 \\
    JNLP & WT+L & 0.6094 & 5 \\
    \bottomrule
    \end{tabular}
    \label{tab:task2}
\end{table}

As shown in Table~\ref{tab:task2}, our second run, which utilizes the asymmetric truncation during BERT fine-tuning, achieves the best performance among our submitted three runs and the improvement is nontrivial. This result indicates the effectiveness of the asymmetric truncation. To be specific, the asymmetric truncation attempts to maintain the information in the decision fragment when constructing the input pairs. As an analogy, traditional retrieval models (e.g., BM25) tend to focus more on query items and thus we believe that the decision fragment (also denoted as ``query'') should be paid more attention to in this task. Unfortunately, our other two runs seem to be less effective in this task and the combination of different models does not help significantly. We interpret it from several perspectives. For one thing, considering the limited data scale, fine-tuning for 4 epochs might cause over-fitting. Since the division of training and validation set is constant in our experiments, the potential bias in the data division might mislead the model selection. For another, according to the task definition, a relevant paragraph does not necessarily entail the query fragment and thus the semantic understanding technique is required more than the exact matching one, which might explain why the combination does not work well in this task. Generally, our best run ranks 4th among all of the submitted runs in this task. The top-ranked one (``BMWT'' from the ``JNLP'' team) does outperform all of the other runs considerably and is worth further discussion. 

\section{Conclusion and Future Work}
In this paper, we introduce our methods in two case law tasks in COLIEE-2020, including a legal case retrieval task and a legal case entailment task. We investigate both semantic understanding and exacting matching methods and contribute three distinct runs in each task, respectively. In the legal case retrieval task, compared between the two types of models, the neural model that focuses on semantic understanding (BERT-PLI) outperforms the traditional retrieval model that is based on exact matching (Duet) a lot. Moreover, the combination of two types of methods enhances each single one, which suggests that semantic understanding models and exact matching ones are complementary in this task. We rank 2nd in Task 1 and our best run ranks 3rd, which achieves rather close performance to the top-ranked ones. However, in the legal case entailment task, the semantic understanding ability is more important. We also find that reducing the information loss in the query fragment can help a lot in this task. In task 2, our best run ranks 4th among all runs and there still exists a performance gap from the top-1 run. 

As an attempt to tackle the challenges in legal case retrieval and entailment, there still exist some limitations in our work that we would like to list as future directions. In this paper, different models are combined simply through RankSVM, a common learning to rank algorithm, while other more sophisticated mechanisms, such as pseudo-relevant feedback, are worth further investigation. Besides, the bias in the dataset division would mislead the optimization process of the supervised learning models. Cross-validation might be an alternative way in this scenario. Last but not least, we employ the BERT pre-trained on the common text by \textit{Google} and merely fine-tune it on the limited legal data. With more available legal documents, pre-training a BERT in the legal domain might enhance most of the BERT-based models in this scenario. 

%
%
\bibliographystyle{splncs04}
\bibliography{reference}
\end{document}